\documentclass{desyproc}
\usepackage{graphicx}
\usepackage{caption}
\usepackage{subcaption}
\pdfoutput=1

\begin{document}
\title{Low-$x$ QCD from CMS}

\author{{\slshape Maciej Misiura}\\[1ex]
University of Warsaw, Faculty of Physics, \\Hoza 69, 00-681 Warsaw, Poland\\
\\
on behalf of CMS Collaboration
 }

\contribID{misiura\_maciej}


\acronym{EDS'09} 

\maketitle

\begin{abstract}
In this paper, selected CMS measurements sensitive to low-x QCD are presented: inclusive cross-sections for forward-central jets  production  at $\sqrt{s}= 7\mathrm{\ TeV}$, inclusive cross-sections for forward jets  production at 7 TeV  and 8 TeV, inclusive to exclusive dijets production cross-sections ratios and correlations of Mueller-Navelet dijets at 7 TeV. Results for data are compared to predictions of theoretical  models. 
\end{abstract}

\section{Introduction}

The term ãsmall-$x$Ó refers to interactions in which a small fraction of the proton momentum is carried by an interacting parton. The Compact Muon Solenoid (CMS) experiment at the CERN Large Hadron Collider (LHC) provides valuable testing ground for the QCD in the small-$x$ region where searches for signs of BFKL evolution can be performed. In this paper four measurements sensitive to low-$x$ processes are described: (1) measurement of inclusive  cross-sections for forward jets production with associated production of central jets with data taken at $\sqrt{s}= 7\mathrm{\ TeV}$, (2) measurement of inclusive  cross-sections for forward jets with data taken at $\sqrt{s}= 8\mathrm{\ TeV}$, (3) measurement of inclusive to exclusive dijets production cross-section ratios at $\sqrt{s}= 7\mathrm{\ TeV}$ and (4) measurement of angular correlation of pairs of jets widely separated in rapidity - Mueller Navelet dijets - at $\sqrt{s}= 7\mathrm{\ TeV}$.

A detailed description of the CMS detector can be found in~\cite{cms}.  The most crucial subdetectors for the presented analyses are the calorimeters, especially the Hadronic Forward (HF) calorimeter that covers pseudorapidity region from $|\eta|=3$ to $|\eta|=5.2$. 
 
\section{Cross-sections measurements}

\begin{figure}
   \centering
\begin{subfigure}{0.39\textwidth}
                \centering
                \includegraphics*[width=\textwidth]{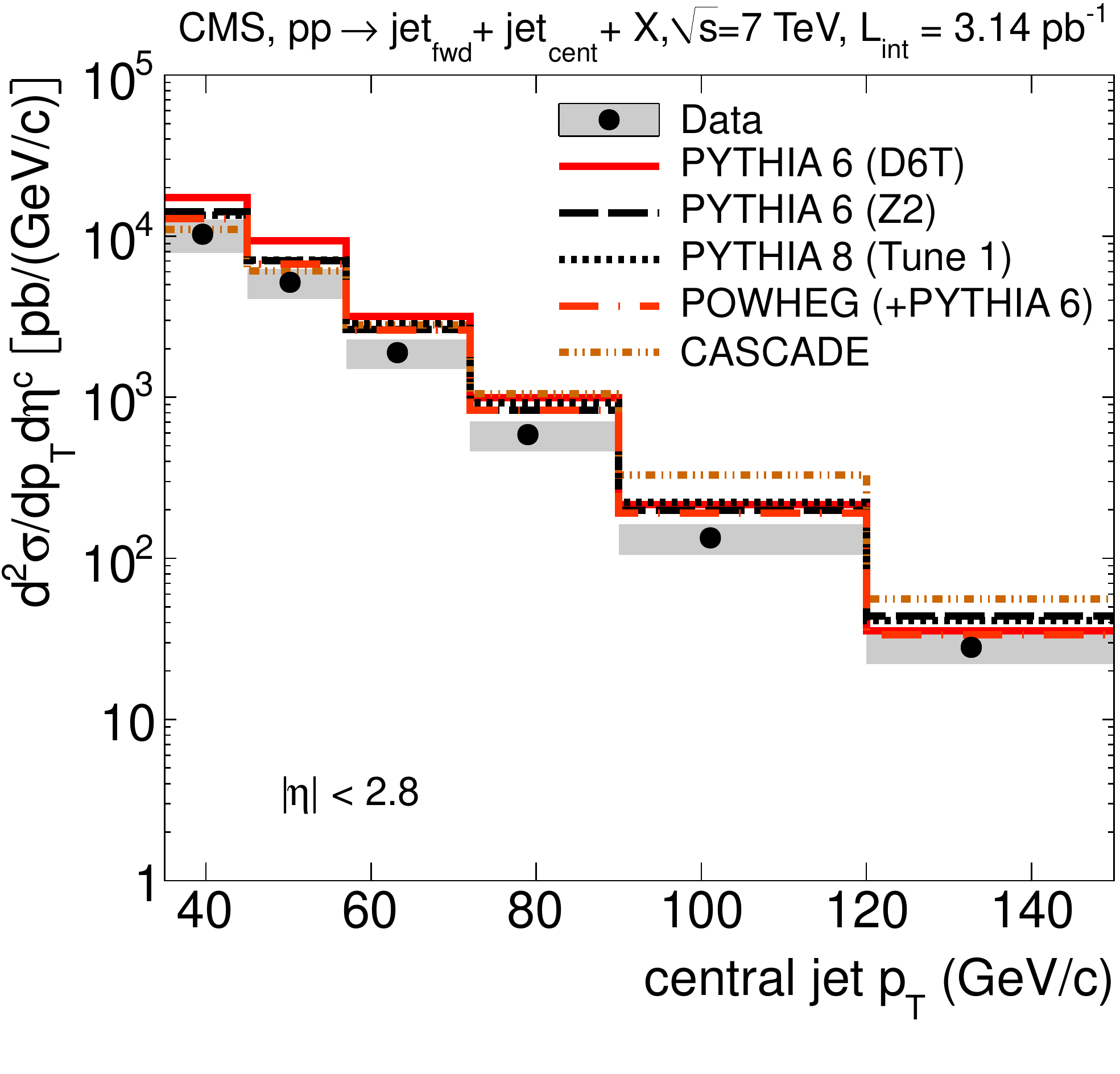}
\end{subfigure}
\begin{subfigure}{0.43\textwidth}
                \centering
                \includegraphics*[width=\textwidth]{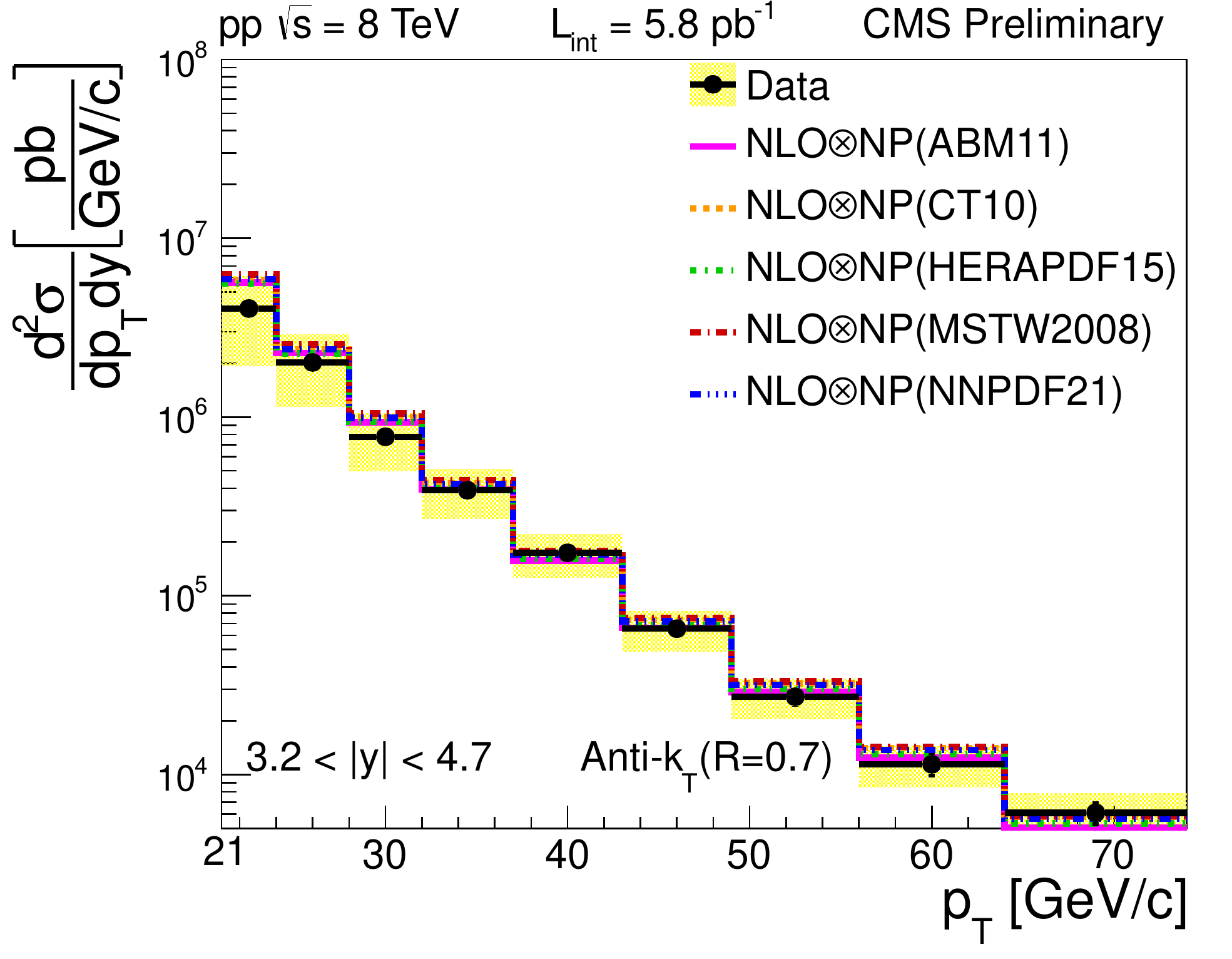}
\end{subfigure}
\caption{Results for the measurements of inclusive cross-section for forward-central dijet as a function of central jet $p_T$ (top left) with data taken at 7 TeV. Results for the measurements of inclusive cross-section for forward jets with $p_T>21\mathrm{\ GeV}$ (right) taken at 8 TeV.}
\label{fig:xsec}
\end{figure}

The aim of the analysis presented in~\cite{incl_7TeV} was the measurement of the differential inclusive cross-sections for two topologies: forward jets ($3.2<|\eta|<4.7$) and forward jet associated with jet emitted in the central part of the detector ($|\eta|<2.8$). Cross-sections were measured as a function of the jet transverse momentum $p_{T}$ taking into account only jets with $p_{T}>35~\mathrm{GeV}$. Data was taken in 2010 with the collision energy $\sqrt{s}= 7\mathrm{\ TeV}$. Results for the data were unfolded to the stable particle level. The dominating systematical uncertainty comes from the  Jet Energy Scale. Results for the data are compared to predictions of various Monte Carlo generators and to NLO calculations. Detailed results can be found in~\cite{incl_7TeV}. The forward jet spectrum is consistent within uncertainties with predictions of theoretical models with and without associated production of a central jet. Results for the forward-central dijet spectrum are presented in the Figure~\ref{fig:xsec} (left) as a function of central jet $p_{T}$. The models studied tend to predict larger values of cross-sections than observed in the data. The best predictions are provided by Herwig 6 and Herwig++ (shown in~\cite{incl_7TeV}). 

In the paper~\cite{incl_8TeV} the $p_T$ spectrum of forward jets up to pseudorapidity $|\eta|<4.7$ is presented. The analysis is based on 2012 data taken at $\sqrt{s}= 8\mathrm{\ TeV}$ during a dedicated low pile-up run. The measured $p_{T}$ range is 21~GeV to 75~GeV. Results were unfolded to the stable particle level. The dominating source of systematic uncertainty was Jet Energy Scale. Results are compared to NLO predictions with corrections for non-perturbative effects. In the Figure~\ref{fig:xsec} (right) results for the most forward rapidity bin ($3.2<|\eta|<4.7$) are presented. Theoretical predictions overestimate the measured cross-sections.

\section{Mueller-Navelet dijets measurements}

\begin{figure}
   \centering
\begin{subfigure}{0.37\textwidth}
                \centering
                \includegraphics*[width=\textwidth]{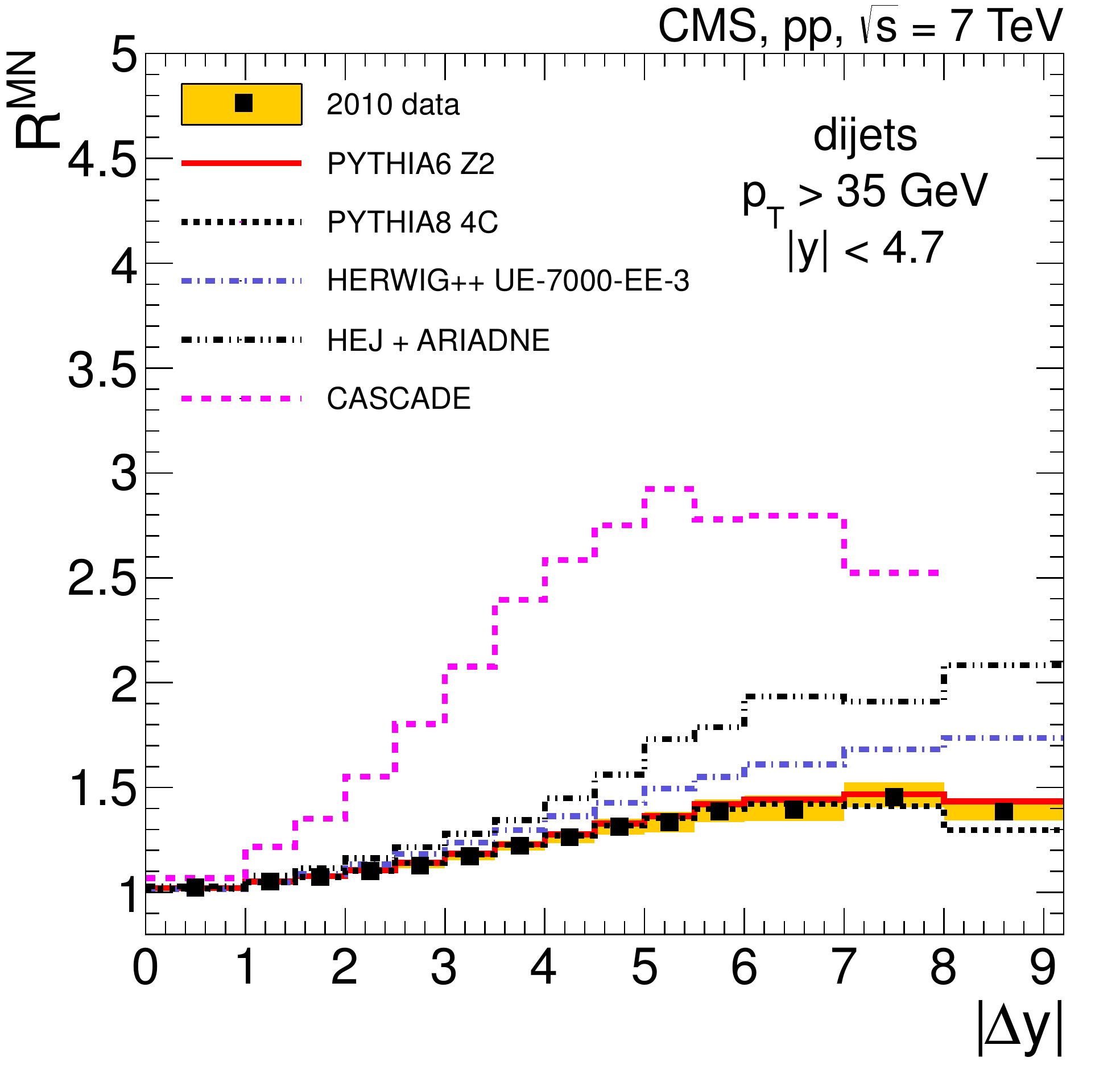}
\end{subfigure}
\begin{subfigure}{0.37\textwidth}
                \centering
                \includegraphics*[width=\textwidth,angle=90]{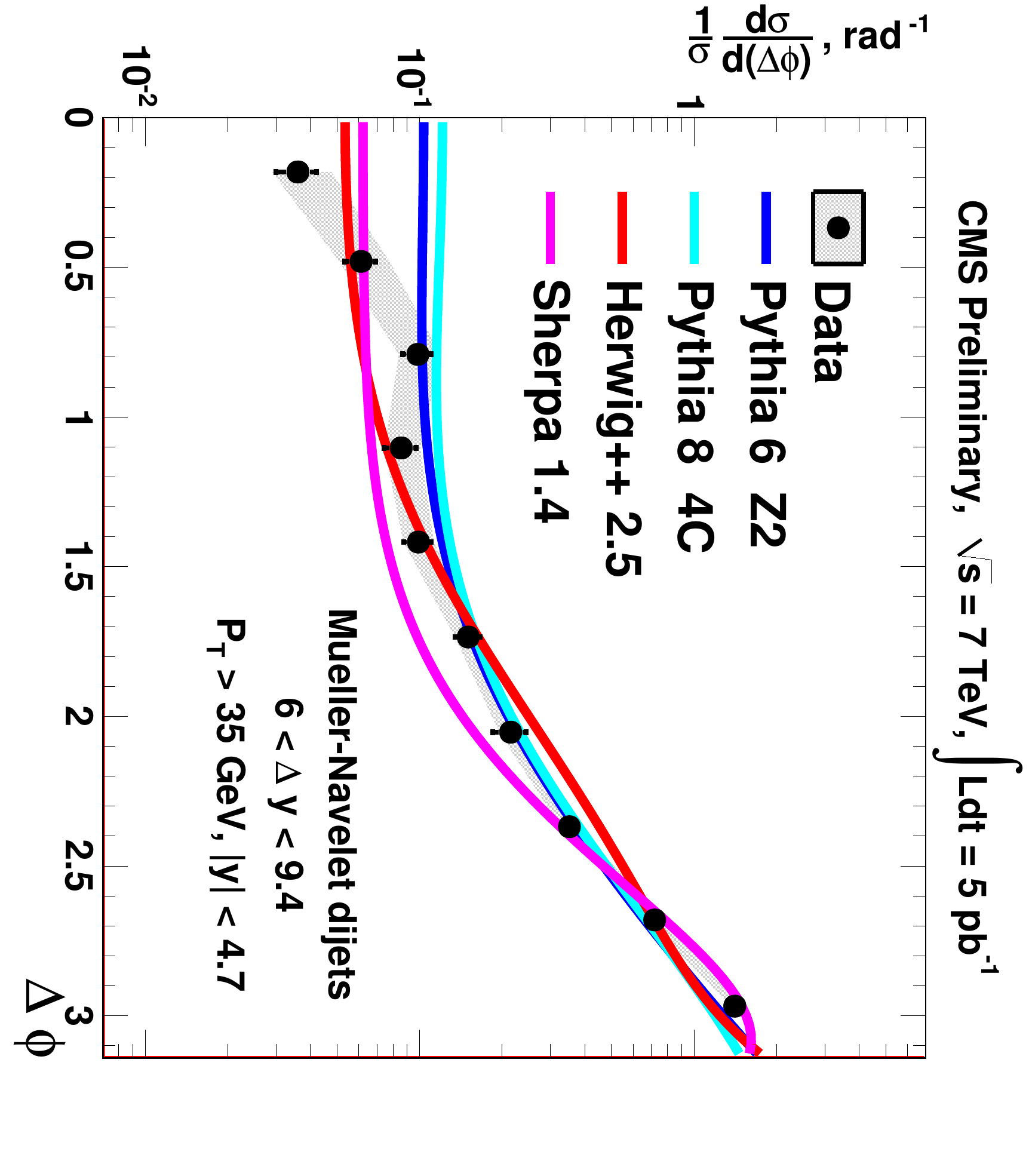}
\end{subfigure}
\caption{Results for the measurements of $R^{\mathrm{MN}}$ as a function of $\Delta\eta$ (left). Distribution of azimuthal angle difference for most forward (right) $\Delta\eta$ bin in MN pairs analysis. }
\label{fig:MN}
\end{figure}

In the analysis presented in \cite{kfac} ratios of cross-sections for dijet production are measured: $R^{\mathrm{incl}}=\sigma_{\mathrm{incl}}/\sigma_{\mathrm{excl}}$ and $R^{\mathrm{MN}}=\sigma_{\mathrm{MN}}/\sigma_{\mathrm{excl}}$ as a function of $\Delta\eta$. Only jets with $p_{T}>35~\mathrm{GeV}$ and $|\eta|<4.7$ are considered. The inclusive cross section $\sigma_{\mathrm{incl}}$ is obtained by taking all pairwise combinations of jets in the event. The exclusive cross section $\sigma_{\mathrm{excl}}$ is measured from a subsample of events containing only one pair of jets. Mueller-Navelet sample ($\sigma_{\mathrm{MN}}$) is defined, taking from combinations of jet pairs the one with the largest $\Delta\eta$ separation.
The data were collected in 2010 at 7~TeV. The results are corrected to the stable particle level. The $R^{\mathrm{MN}}$ results are presented in the Figure~\ref{fig:MN} (left). Pythia 6 tune Z2 and Pythia8 tune 4C agree with the measurement within the systematic uncertainty, represented as yellow band. Predictions of Herwig++ and HEJ (+Ariadne) are larger than observed. The discrepencies become larger with increasing separation in pseudorapidity. CASCADE, in which elements of BFKL approach in the LL approximation are implemented, predicts ratios much larger than observed in the data. Results of $R^{\mathrm{incl}}$ and further details can be found in~\cite{kfac}.

In~\cite{MN} angular correlations of Mueller-Navelet dijets as a function of $\Delta\eta$ are presented. The analysis is based on 2010 data collected at 7 TeV. Jets with $p_{T}>35~\mathrm{GeV}$ and $|\eta|<4.7$ are selected. The results are compared to both DGLAP and BFKL-based MC generators, and to NLL BFKL calculations. For each MN pair the angular distance is calculated: $\Delta\phi=\phi_1-\phi_2$. Not only $\Delta\phi$ distributions are studied, but also the average cosines: $C_n=\left < \cos \left ( n\left(\Delta\phi - \pi\right) \right ) \right >$ for $n\in\{1,2,3\}$, corresponding to the coefficients of a Fourier series in $\Delta\phi$, and their ratios. In the Figure~\ref{fig:MN} (right) $\Delta\phi$ distributions for the bin with the largest $\Delta\eta$ is presented. For low $\Delta\phi$ DGLAP-based MCs show deviation from the data. Considering the average cosines (not shown here) at mid and high rapidity description of data by DGLAP predictions is worse. On the other hand the CASCADE generator, implementing elements of the BFKL approach, does not provide description of data in full $\Delta\eta$ range. The NLL BFKL calculations provide a good description of $C_n$ ratios, nevertheless they are predicted with large theoretical uncertainties (see~\cite{MN}).

\section{Summary}

Four measurements in the low-$x$ region of the phase space have been presented. In the jet measurements discussed here there is no clear indication for the presence of BFKL effects in the data. There are some discrepancies between predictions and the data that should be further studied.


\begin{footnotesize}

\end{footnotesize}
\end{document}